\documentstyle[aps,prd,preprint,epsfig,tighten]{revtex}
\begin{document}
%..........................................................................
\preprint{\makebox{\begin{tabular}{r}
							BARI-TH/343-98	\\
                                                                        \\
\end{tabular}}}
%..........................................................................
\draft
\title{Comments on atmospheric neutrino oscillation scenarios \\
	with large $\nu_\mu\leftrightarrow\nu_e$ transitions}
\author{ G.L.\ Fogli, E.\ Lisi, A.\ Marrone, and G.\ Scioscia }
\address{Dipartimento di Fisica and Sezione INFN di Bari\\
             Via Amendola 173, I-70126 Bari, Italy \\ }
\date{\today}
\maketitle
\begin{abstract}%............................................................
The evidence for $\nu_\mu$ disappearance in the Super-Kamiokande atmospheric
neutrino experiment and the negative searches for $\nu_e$ disappearance
in the CHOOZ reactor experiment can be easily reconciled by assuming
oscillations with large amplitude in the $\nu_\mu\leftrightarrow\nu_\tau$
channel and small (or null) amplitude in the $\nu_\mu\leftrightarrow\nu_e$
channel. It has been claimed, however, that some oscillation
scenarios with large $\nu_\mu\leftrightarrow\nu_e$ mixing can also be
constructed
in agreement with the present data. We investigate
quantitatively two such scenarios: 
(a) threefold maximal mixing; and (b) attempts
to fit 
all sources of evidence for oscillations (solar, atmospheric, and
accelerator) with three neutrinos. 
By using mainly Super-Kamiokande data, we show that the case
(a) is disfavored, and that the case (b) is definitely ruled out.
\end{abstract}%.............................................................
\medskip
\pacs{PACS: 14.60.Pq, 13.15.+g, 95.85.Ry}

%%%%%%%%%%%%%%%%%%%%%%%%%%%%%%%%%%%%%%%%%%%%%%%%%%%%%%%%%%%%%%%%%%%%%%%%%%%%
\section{Introduction}
%%%%%%%%%%%%%%%%%%%%%%%%%%%%%%%%%%%%%%%%%%%%%%%%%%%%%%%%%%%%%%%%%%%%%%%%%%%%

As well known, the evidence for atmospheric $\nu_\mu$ disappearance
observed in the Super-Kamiokande (SK) experiment \cite{To98} can be 
explained by two-family oscillations with large mixing in the
$\nu_\mu\leftrightarrow\nu_\tau$ 
channel \cite{SK98}.
When this solution, known as twofold maximal mixing, is assumed,
the relevant flavor oscillation probabilities 
$P_{\alpha\beta}$ read: 
\begin{eqnarray}
P_{ee} &=& 1\ ,\\
P_{\mu e} &=& 0\ ,\\
P_{\mu\mu} &=& 1-\sin^2(\Delta m^2_{\rm atm} L/4E)\ ,
\end{eqnarray}
where $L$ and $E$ are the neutrino pathlength and energy, respectively, and
the  experimentally inferred value of the
neutrino squared mass difference  is
\cite{Sc99,Fo99}
%...............................................................
\begin{equation}
\Delta m^2_{\rm atm} \sim 3 \times 10^{-3} {\rm \ eV}^2\ 
\label{deltamatm}
\end{equation}
%...............................................................
within a factor of about two. 

Oscillations in the
$\nu_\mu\leftrightarrow\nu_\tau$ channel are not affected by matter effects,
and are consistent with the negative results of the CHOOZ $\nu_e$
disappearance experiment
\cite{Ap98,Ni98}, which requires $P_{ee} \sim 1$ for
$\Delta m^2 \gtrsim 10^{-3}$ eV$^2$.
Conversely, two-family oscillations in the 
$\nu_\mu\leftrightarrow\nu_e$ channel are heavily
affected by matter effects,
do not provide a good fit to SK data \cite{Sc99,Fo99}, and are
independently excluded  by CHOOZ \cite{Ap98}.

Going beyond pure $\nu_\mu\leftrightarrow\nu_\tau$ oscillations,
one should consider the possibility that
the flavor eigenstates $(\nu_e,\nu_\mu,\nu_\tau)$  are linear combinations
of three states
$(\nu_1,\nu_2,\,\nu_3)$ with definite masses
$m_1\leq m_2\leq m_3$ through a unitary matrix
$U_{\alpha i}$:
\begin{equation}
 U_{\alpha i} = \left(\begin{array}{ccc}
c_\omega c_\phi & 
	s_\omega c_\phi & 
		s_\phi\\
-s_\omega c_\psi - c_\omega s_\psi s_\phi &
	c_\omega c_\psi-s_\omega s_\psi s_\phi &
		s_\psi c_\phi\\
s_\omega s_\psi - c_\omega c_\psi s_\phi &
	-c_\omega s_\psi - s_\omega c_\psi s_\phi &
		c_\psi c_\phi\\
\end{array}\right)
\end{equation}
where $(\omega,\phi,\psi)$ are the mixing angles in the standard
parametrization, and a possible CP violating phase has been neglected.
The
parameter $\Delta m^2_{\rm atm}$ should then be taken
equal to one of the following squared mass differences:
%..........................................................................
\begin{eqnarray}
\delta m^2 &=& m^2_2-m^2_1\ , \\
m^2 &=& m^2_3-m^2_2\ ,
\label{deltas}
\end{eqnarray}
%............................................................................
assuming that $\delta m^2\ll m^2$.

Different choices for  $\Delta m^2_{\rm atm}$ have different implications
for three-flavor scenarios. In particular,  if $\Delta m^2_{\rm atm}$ 
is taken within CHOOZ bounds (i.e., $\gtrsim 10^{-3}$ eV$^2$), then
large $\nu_\mu$ disappearance in SK can be generated only by dominant
$\nu_\mu\leftrightarrow\nu_\tau$ oscillations, while
$\nu_\mu\leftrightarrow\nu_e$ transitions must be small (or absent).
 This is the scenario favored by the quantitative analyses
\cite{Fo99,Wi99}.

On the other hand, for $\Delta m^2_{\rm atm}\lesssim 10^{-3}$ eV$^2$ 
the CHOOZ bounds are not operative and, in principle,
 sizable $\nu_\mu\leftrightarrow\nu_e$ oscillations can occur.
Even though it seems difficult to reconcile SK data with
large $\nu_e$ mixing and relatively small $\Delta m^2_{\rm atm}$
\cite{Fo99,Wi99}, nevertheless, this possibility has been recently
claimed in some 
 phenomenological models, including:
a) ``Threefold maximal
mixing'' of solar and atmospheric neutrinos \cite{Ha99,Fo98}; b) Three-flavor
fits to solar, atmospheric,
and accelerator evidence for neutrino oscillations 
\cite{Th98,Ba98,Co99,Oh99,St99}. 
Such models with large $\nu_e$ mixing 
are representative of the two possible identifications for
$\Delta m^2_{\rm atm}$, which is taken equal to
$m^2$ in model (a) and to $\delta m^2$ in model (b).

Since both models (a) and (b)  have received considerable and continued
attention in the neutrino physics literature%
%---------------------------------------------------------------------------
\footnote{For pre-SK studies see, e.g.,  
Ref.~\protect\cite{Ha97} for case (a),  Ref.~\protect\cite{Ac97} 
for case (b), and
\protect\cite{Fo97} for a general review of three-flavor scenarios.},
%------------------------------------------------------------------------------
it seems useful
to assess clearly 
their phenomenological status through a quantitative
study of the  Super-Kamiokande observations. In this work, we perform such
an analysis and show that model (a) 
is disfavored (although not yet excluded),
while model (b) 
is definitely ruled out. We also try to trace the origin
of different claims by other authors.

%%%%%%%%%%%%%%%%%%%%%%%%%%%%%%%%%%%%%%%%%%%%%%%%%%%%%%%%%%%%%%%%%%%%%%%%%%%%
\section{Threefold maximal mixing}
%%%%%%%%%%%%%%%%%%%%%%%%%%%%%%%%%%%%%%%%%%%%%%%%%%%%%%%%%%%%%%%%%%%%%%%%%%%%

Threefold maximal mixing is
defined by a democratic mixing matrix: $|U^2_{\alpha i}|=1/3$
$(\alpha=e,\mu,\tau;\, i=1,2,3)$, so that
all oscillation channels are open. The parameters $m^2$ and $\delta m^2$
are assumed to drive atmospheric and solar neutrino oscillations, respectively,
so that $\delta m^2\ll m^2$ can be assumed.
The phenomenological implications of
such model have been intensively studied by
Harrison, Perkins, and Scott
(see \cite{Ha99,Ha97} and references therein). 

As far as atmospheric neutrinos are concerned, one can take
$m^2=\Delta m^2_{\rm atm}$ and $\delta m^2\simeq 0$ 
(one mass scale dominance \cite{Fo99}).
Since the mixing matrix elements (squared) are fixed to the
common value 1/3, 
the only parameter probed by SK data is $m^2$.
In the following, we discuss how matter effects and fits to the SK
and CHOOZ data constrain the value of $m^2$ for threefold maximal mixing,
and distinguish this scenario from twofold maximal mixing.

\subsection{Oscillations in matter}

The three-flavor oscillation probabilities 
for the case of $\delta m^2\ll m^2=\Delta m^2_{atm}$
and generic mixing can be found, e.g., in Appendix~C of Ref.~\cite{Li97}.
The specific case of threefold maximal mixing,  obtained by
taking $s^2_\psi=1/2$ and $s^2_\phi=1/3$, leads to the well-known
results for vacuum oscillations:
\begin{eqnarray}
P^{\rm vac}_{ee} &=& 1-\frac{8}{9}\, S\ , \\
P^{\rm vac}_{\mu e} 	&=& \frac{4}{9}\, S\ ,\\
P^{\rm vac}_{\mu\mu}	&=& 1-\frac{8}{9}\, S\ , 
\end{eqnarray}
where $S=\sin^2 (m^2 L/4 E)$.

In the threefold maximal mixing scenario (as well as in any
model with large $\nu_e$ mixing) atmospheric neutrino oscillations
below the horizon are heavily affected by
Earth matter effects \cite{Matt}, as emphasized, e.g.,  in 
\cite{Fo99,Fo98,Li97,Li98,Ma98,Fo95,Pa98,Gi98}, 
and recently realized also in \cite{Ha99}.%
%--------------------------------------------------------------
\footnote{This contrasts with solar
neutrino oscillations, where matter effects are ineffective for
threefold maximal mixing \protect\cite{Ha96}.}
%---------------------------------------------------------------
Since the majority of upgoing leptons in SK are produced
by neutrinos traversing only the Earth mantle, 
the bulk of matter effects
can be understood by assuming 
a constant electron density 
($N_e\simeq 2$ mol/cm$^3$)
 along the neutrino trajectory. 
In this case, the flavor oscillation probabilities $P^{\rm mat}_{\alpha \beta}$
depend on the (constant) matter-induced squared mass term $A$:
\begin{equation}
	A = 2\sqrt{2}\, G_F\, N_e\, E \simeq 0.3 \times 10^{-3}
	\, \frac{E}{\rm GeV}
	\ \ [{\rm eV}^2]\ ,
\end{equation}
and can be computed analytically.
Explicit expressions for $P^{\rm mat}_{\alpha\beta}$ can be found in
\cite{Li97}. Two limit cases are particularly illuminating for
understanding matter effects in threefold maximal mixing: 
low energy events ($A\ll m^2$) and high energy events ($A\gg m^2$).

The regime $A\ll m^2$  corresponds (for $m^2\sim 10^{-3}$ eV$^2$)
to $E\ll 3$ GeV, and is thus  relevant for the
so-called sub-GeV (SG) atmospheric neutrino events. In this limit
it turns out that 
\cite{Li97}
\begin{eqnarray}
P^{\rm mat}_{ee} &=& P^{\rm vac}_{ee}\ , \\
P^{\rm mat}_{\mu e} 	&=& P^{\rm vac}_{\mu e}\ ,\\
P^{\rm mat}_{\mu\mu}	&=& P^{\rm vac}_{\mu \mu}-\delta P\ , 
\end{eqnarray}
where the term $\delta P$, for threefold maximal mixing, is given
by:
\begin{eqnarray}
\delta P &=&   \frac{1}{3}
\sin^2\left( \frac{A  L}{6 E}\right)\\
&\simeq & \frac{1}{3} \sin^2(L/R_\oplus)\\
&\simeq & \frac{1}{3} \sin^2(2\cos\Theta)\ .
\end{eqnarray}
In the above equations,
$\Theta$ is the neutrino 
zenith angle ($L\simeq 2 R_{\oplus}\cos\Theta$),
and the accidental equality $A/6E \simeq R_{\oplus}$ for
$N_e=2$ mol/cm$^3$ has been used.
It thus appears that, 
in the sub-GeV sample,  no significant matter effects
can be expected for electron events, while  the muon event
rate should be further suppressed via the energy-independent
term $\delta P \lesssim 10\%$. This term
is generated by the effective splitting in matter of the two
vacuum-degenerate  eigenstates $(\nu_1,\nu_2)$
\cite{Li97,Li98,Gi98}.

The regime $A\gg m^2$  corresponds (for $m^2\sim 10^{-3}$ eV$^2$)
to $E\gg 3$ GeV, and is thus  relevant for the
so-called upward through-going
muon events (UP$\mu$). In this limit it turns out that
\begin{eqnarray}
P_{ee}^{\rm mat} &=& 1\ ,\\
P_{\mu e}^{\rm mat} &=& 0\ ,\\
P_{\mu\mu}^{\rm mat} &=& 1-\sin^2\left(\frac{2}{3} m^2
\frac{L}{4E} \right)\ ,
\end{eqnarray}
namely, threefold maximal mixing in matter becomes
equivalent to twofold maximal mixing in vacuum
\cite{Fo98,Pa98}, but with an
effective squared mass difference decreased by a factor 2/3: 
$m^2_{\rm mat}=2/3\,m^2$ \cite{Ha99,Fo98}.

The above approximations are useful for
a qualitative understanding of oscillations in matter, but
cannot be applied in the regime of
intermediate energies [$E\sim O(3$ GeV)], 
typical of the so-called
multi-GeV (MG)  events. Therefore, for 
numerical calculations we prefer to solve 
exactly, at any energy, the neutrino propagation 
equations along the Earth density
profile. Details of our computation technique can be found 
in our previous works \cite{Fo99,Wi99,Li97,Fo95}; see, in particular,
\cite{Fo99} for SK observables. 
Representative results 
of matter effects
for threefold maximal mixing are shown in Fig.~1.

Figure~1 shows the SK distributions of sub-GeV electrons and muons
(SG$e$ and SG$\mu$), of multi-GeV electrons and muons (MG$e$ and MG$\mu$)
and of upward-going muons (UP$\mu$), as a function of the lepton
zenith angle $\theta$, for $m^2=10^{-3}$ eV$^2$. 
In each bin, the lepton rates are normalized
to their expectations in the absence of oscillations. The SK data (dots
with error bars)
refer to 45 kTy \cite{Sc99}. The solid histograms (blue in color) 
represent the predictions of threefold maximal mixing including
matter effects; the dashed histograms (red in color)
are obtained by  assuming pure
vacuum oscillations. As expected from the previous discussion, matter effects
are negligible for
the SG$e$ distribution, while they generate an additional
suppression for the SG$\mu$ distribution. The UP$\mu$ distribution
is less ``tilted'' in the presence of matter oscillations, as a result
of the smaller effective squared mass at high energy:
$m^2_{\rm mat}\simeq
2/3\,  m^2$.
At ``intermediate energies'' (MG events), both
the $e$ and $\mu$ distributions are 
affected by oscillations in matter. It can be seen that matter
effects help the fit to SG and MG data, but slightly worsen 
the fit to UP$\mu$ data, as compared to pure vacuum oscillations.

\subsection{Data analysis}

It has recently been claimed that, including matter effects,
threefold maximal mixing with $m^2\sim 10^{-3}$ 
represents a good fit to all the SK data \cite{Ha99,Fo98,Sr99}. 
Although matter effects certainly help to fit the data in
such model, the statements in  \cite{Ha99,Fo98,Sr99}
seem too optimistic as compared with our detailed 
analysis of either 33 kTy \cite{Fo99} or
45 kTy \cite{Wi99} SK data, which disfavors scenarios with
large values of $|U^2_{e3}|$, including
threefold maximal mixing ($|U^2_{e3}|=1/3$). In this section, we clarify and 
further corroborate our previous results,
and argue about 
the differences with the more optimistic claims in \cite{Ha99,Fo98,Sr99}.

Figure~2 shows the SK distributions for threefold maximal mixing
($m^2=10^{-3}$ eV$^2$,  matter effects included), as compared
with the case of twofold maximal mixing $\nu_\mu\leftrightarrow\nu_\tau$
(at $m^2=1$ and $3\times 10^{-3}$ eV$^2$). 
The differences among 
$2\nu$ and $3\nu$ distributions are
relatively small for 
SG$e$, SG$\mu$, and MG$e$ events, while they are more significant for
MG$\mu$ and UP$\mu$ events.  Let us thus focus on
the latter two samples.
For MG$\mu$,
threefold maximal mixing overestimates the bin rate around the horizon
$(\cos\theta\in[-0.2,0.2])$, as compared to twofold maximal
mixing. This effect 
has two components: (i) the relatively low value
of $m^2$ implied by CHOOZ for $3\nu$ maximal mixing
($\lesssim 10^{-3}$ eV$^2$), with respect to the best-fit value
for $2\nu$ maximal mixing $(\sim 3\times 10^{-3})$ eV$^2$; 
and (ii) the further ``matter'' suppression of $m^2$ by a factor
2/3 for the highest-energy part of the
MG$\mu$ sample (partially contained events). 
Both components 
tend to increase the typical oscillation wavelength for $3\nu$ maximal mixing
(as compared
to the $2\nu$ case), so that longer pathlengths are required
to get effective muon suppression.%
%
%----------------------------------------------------------------------
\footnote{The recent ten-bin MG$\mu$ data from SK rapidly drop   
at the horizon, and show no evidence (within errors) for a wavelength 
significantly longer
than for the pure $2\nu$ best-fit case at
$m^2\sim 3\times 10^{-3}$ eV$^2$; see Fig.~3 in 
\protect\cite{Sc99}.}
%----------------------------------------------------------------------
Concerning UP$\mu$'s one faces similar problems. 
Threefold maximal mixing
is unable to fit the slope suggested by the data pattern, both because
$m^2$ is necessarily low, and because of the further effective 2/3
suppression of $m^2$ in matter. For UP$\mu$ events, threefold
maximal mixing with $m^2=10^{-3}$ eV$^2$ is phenomenologically
equivalent to twofold maximal mixing with $m^2=0.67\times 10^{-3}$
eV$^2$---a value which is definitely below the 99\% C.L. range 
allowed by UP$\mu$ data alone (see Fig.~9 in \cite{Sc99}).
 Summarizing, Fig.~2
shows that $2\nu$ and $3\nu$ maximal mixing are discriminated mainly
by MG$\mu$ and UP$\mu$ data; the low value of $m^2$ allowed by CHOOZ
for $3\nu$ maximal mixing ($\lesssim 10^{-3}$ eV$^2$) prevents
a good fit to the 
MG$\mu$ rate around the horizon and to the slope of UP$\mu$'s,
in contrast to the very good fit provided
with $2\nu$ maximal mixing at $m^2\sim 10^{-3}$ eV$^2$.

Figure~3 shows $\chi^2$-fits 
to all the SK data (SG+MG+UP) for twofold and threefold
maximal mixing, as a function of $m^2$. Details of our statistical
analysis can be found in \cite{Fo99}. The dot-dashed line 
(black in color) refers to the
twofold maximal mixing, characterized by $\chi^2_{\rm min}\simeq 20$
and $m^2\simeq 2.8 \times 10^{-3}$ eV$^2$ at best fit. The allowed
range of $m^2$ is in very good agreement with the SK official analysis
(compare with Fig.~10 in \cite{Sc99}). The dashed line (red in color) 
refers to threefold maximal mixing (SK data only), which in principle
could provide a good fit for $m^2\simeq 4\times 10^{-3}$ eV$^2$, if
such value were not excluded by CHOOZ.  Adding the CHOOZ constraint
(solid line, blue in color), the best fit is pushed to lower $m^2$
values and the value of $\chi^2_{\rm min}$ increases up to $\sim 35$.
Although this is not high enough to be ruled out, 
it is definitely worse than in the two-flavor
case since, as 
observed in Fig.~2, $3\nu$ maximal mixing fails to fit the 
UP$\mu$ and horizontal MG$\mu$ data. The situation would be even 
worse if matter effects were hypothetically switched off (dotted 
line, green in color).

The quantitative results of Fig.~3 indicate that the two-flavor
and three-flavor scenarios with maximal mixing differ by 
$\Delta \chi^2\sim 15$; therefore, if $3\nu$ mixing is left unconstrained,
the minimum will fall close to the $2\nu$ case with zero
$\nu_e$ mixing, while $3\nu$ cases 
with large $\nu_e$ mixing will be
disfavored. This is indeed the 
pattern found in the $3\nu$ parameter space analysis
of \cite{Fo99,Wi99}. 
However, it should also be said that the $3\nu$ maximal mixing
fit to the SK data is not terribly bad, so it is
wiser to wait for 
higher statistic in the MG$\mu$ and UP$\mu$ samples, or for more 
stringent reactor bounds on $m^2$, before (dis)proving it definitely.

We conclude by examining possible reasons for the more optimistic
results found in \cite{Ha99,Fo98,Sr99}. The work \cite{Sr99} did not
include UP$\mu$ data, which contribute to worsen the fit for threefold
maximal mixing. The work \cite{Fo98}, moreover, considered only
the up-down asymmetry for SG and MG data, thus excluding also the important
information given by nearly horizontal MG$\mu$. The comparison
with \cite{Ha99} is more delicate since, in principle, a data set
similar to ours is analyzed. However, the lack of details and of explicit
comparison with SK calculations in \cite{Ha99} prevents us to trace
the source of the differences. Nevertheless, a comparison of our
Fig.~3 with the corresponding results in \cite{Ha99} (see their 
Fig.~7) 
gives us some hints. It is quite evident that, for the two-flavor oscillation
case, the preferred range for $m^2$ in \cite{Ha99}
is biased towards low values,
as compared with our range in Fig.~3
and with the SK official analysis 
(Fig.~10 in \cite{Sc99}). This is surprising, since the authors
of \cite{Ha99} include also the old Kamiokande data, 
which should rather
pull the fit towards slightly higher $m^2$ values. A bias towards
low $m^2$'s artificially increases the chance that
threefold maximal
mixing can survive below CHOOZ bounds.

In conclusion, we find that 
threefold maximal mixing is significantly disfavored
with respect to twofold maximal mixing 
$(\Delta \chi^2_{\rm min }\simeq 15 )$. The most discriminating
data samples are MG$\mu$ and UP$\mu$. However, further reactor
data, and atmospheric neutrino data in the 
intermediate and high energy range, are needed
in order to (dis)prove the model with
higher confidence. The Super-Kamiokande collaboration itself has now all
the tools to investigate quantitatively generic three-flavor
 scenarios including matter effects
\cite{Ok99},
so as
to settle definitely some differences existing among present
independent phenomenological analyses.

%%%%%%%%%%%%%%%%%%%%%%%%%%%%%%%%%%%%%%%%%%%%%%%%%%%%%%%%%%%%%%%%%%%%%%%%%%%%
\section{Solar~+~atmospheric~+~LSND scenarios}
%%%%%%%%%%%%%%%%%%%%%%%%%%%%%%%%%%%%%%%%%%%%%%%%%%%%%%%%%%%%%%%%%%%%%%%%%%%%

The sources
of evidence for neutrino oscillations coming from solar and atmospheric
neutrinos, as well as from the Liquid Scintillator Neutrino Detector
experiment (LSND) \cite{LSND}
suggest three widely different neutrino mass square differences.
Therefore, they cannot be reconciled in a single three-flavor scenario,
unless some data are ``sacrificed'' \cite{Fo97}. One possibility
\cite{Ac97}, indicated
as ``case 3c'' in Table~VI of Ref.~\cite{Fo97}, has been recently revived 
in several works \cite{Th98,Ba98,Co99,Oh99,St99} which claim that
it can provide a good fit to
the SK atmospheric neutrino data, and also explain most of the 
solar neutrino deficit and the LSND data.
In this section, we briefly discuss such scenario, and show quantitatively that
it is ruled out by the SK atmospheric $\nu$ data.

The starting hypothesis is that $(\delta m^2,m^2)=(\Delta m^2_{\rm atm},
\Delta m^2_{\rm LSND})$, so that: (i) concerning solar neutrinos,
both $\delta m^2$ and $m^2$ drive energy-averaged oscillations, which
can explain the bulk of the solar neutrino deficit but not possible
distortions in the energy spectrum; (ii) for atmospheric neutrinos,
$\delta m^2$ drives the energy-dependent oscillations while $m^2$
gives an energy-averaged contribution; and (iii) for LSND,
$\delta m^2$ can be taken effectively equal to zero, and oscillations
are driven by $m^2\sim O({\rm 1\ eV}^2)$.
 In order to get the desired 
$P_{\mu e}^{\rm LSND}=4U^2_{e3}U^2_{\mu 3}\sin^2(m^2 L/4E)\sim{\rm\ few\
per\ mill}$, the heavy state $\nu_3$ is taken close
to $\nu_\tau$ ($U^2_{\tau 3}\sim 1$), so that $(\nu_1,\nu_2)$ are
basically linear combinations of $(\nu_\mu,\nu_e)$. By choosing nearly
maximal mixing between $\nu_{1,2}$ and $\nu_{e,\mu}$, one then hopes to
solve both the atmospheric neutrino anomaly and the solar neutrino
deficit through dominant $\nu_\mu\leftrightarrow \nu_e$ oscillations
driven by $\delta m^2=\Delta m^2_{\rm atm}$. The CHOOZ results then constrain
$\delta m^2$ to range below $\sim 10^{-3}$ eV$^2$.

The mixing matrix for such scenario
is rather strongly constrained by the requirement to
fit solar, atmospheric, and LSND data. Indeed,
the phenomenological matrices found in different works 
\cite{Th98,Ba98,Co99,Oh99,St99}
are similar to each other, with small differences which are not important
for our discussion. For definiteness, we report the matrix $U_{\alpha i}$
found in the paper \cite{Th98} by Thun and McKee:
\begin{equation}
U_{\alpha i} \simeq \left(
\begin{array}{ccc}
0.78 & 0.60 & 0.18\\
-0.61 & 0.66 & 0.44\\
0.15 & -0.45 & 0.88
\end{array}
\right)\ .
\end{equation}
In the same work \cite{Th98}, the best-fit value of $m^2$ is 0.4 eV$^2$.
Such choice for the mass-mixing parameters will be dubbed ``the Thun-McKee
scenario.''

Figure~4 shows the expected zenith distributions of lepton events in SK
for the Thun-McKee scenario at $\delta m^2=10^{-3}$ eV$^2$, with and without
matter effects. Matter effects play an important role in suppressing 
oscillations: when the $(\nu_\mu,\nu_e)$ mixing is (nearly) maximal
in vacuum, it can only be smaller in matter. A similar pattern, in fact,
is found for pure two-flavor $\nu_\mu\leftrightarrow\nu_e$ oscillations
(see Fig.~9 of \cite{Fo99}). It can be seen in Fig.~4 that the Thun-McKee
scenario is ruled out by the SK data for several reasons: 
(i) the SG$e$ and MG$e$ upgoing event rates are overestimated;
(ii) the SG$\mu$ and MG$\mu$ up-down asymmetries are largely underestimated;
(iii) the predicted UP$\mu$ rate is too low; and (iv) matter effects
play an important role in worsening the fit.%
%---------------------------------
\footnote{Notice that the works \protect\cite{Th98,Ba98,Co99,Oh99,St99}
do not include matter effects, adopt approximate estimates of the
SK event rates, and basically
use only a subset of the data, namely,
the $\mu/e$ ratio of contained events.}
%--------------------------------------
The situation is not
improved by lowering the value of $m^2$, as it can be seen in Fig.~5.
We find that the Thun-McKee scenario, in any case, gives
$\chi^2\gtrsim 100$, and is thus definitely ruled out by the SK data
{\em alone}.%
\footnote{The situation is similar for the slightly different
mass-mixing parameter choices 
performed in \cite{Ba98,Co99,Oh99,St99}.}

In addition, the Thun-McKee scenario does not
provide a good fit to the current reactor and accelerator oscillation
data, including LSND, as already observed
in \cite{Fo97}. In fact, the quantitative three-flavor analysis of
laboratory data
performed in \cite{Sc97} selects only two possible solutions at
$90\%$ C.L., namely, $\nu_3\sim\nu_e$ or $\nu_3\sim\nu_\mu$,
while $\nu_3\sim\nu_\tau$ is highly disfavored.

In conclusion, we find that
the Thun-McKee scenario is not a viable explanation
of the solar+atmospheric+LSND data. The SK data are sufficient to rule it out
definitely.
The much more optimistic claims in  \cite{Th98,Ba98,Co99,Oh99,St99} 
are not substantiated by quantitative  calculations
of electron and muon event rates in SK.

%%%%%%%%%%%%%%%%%%%%%%%%%%%%%%%%%%%%%%%%%%%%%%%%%%%%%%%%%%%%%%%%%%%%%%%%%%%%
\section{Conclusions}
%%%%%%%%%%%%%%%%%%%%%%%%%%%%%%%%%%%%%%%%%%%%%%%%%%%%%%%%%%%%%%%%%%%%%%%%%%%%

We have made a detailed comparison of the SK atmospheric $\nu$ data
with the predictions of 
two popular
models characterized by large $\nu_\mu\leftrightarrow\nu_e$
transitions: (a)
Threefold maximal
mixing \cite{Ha99,Fo98}; b) Three-flavor
fits
to solar, atmospheric,
and LSND data 
\cite{Th98,Ba98,Co99,Oh99,St99}. 
We have found that
model (a) \cite{Ha99} 
is disfavored (although not yet excluded),
while model (b) \cite{Th98,Ba98,Co99,Oh99,St99} 
is definitely ruled out. The origin
of different claims by other authors has been elucidated.

%%%%%%%%%%%%%%%%%%%%%%%%%%%%%%%%%%%%%%%%%%%%%%%%%%%%%%%%%%%%%%%%%%%%%%%%%%%%
\acknowledgments

GLF and EL thank the organizers of the Neutrino Oscillation Workshop
(Amsterdam) and of the
School of Physics ``Bruno Pontecorvo'' (Capri), respectively, where
preliminary results of this work were presented.

%%%%%%%%%%%%%%%%%%%%%%%%%%%%%%%%%%%%%%%%%%%%%%%%%%%%%%%%%%%%%%%%%%%%%%%%%%%%%%%
% 			R E F E R E N C E S 
%%%%%%%%%%%%%%%%%%%%%%%%%%%%%%%%%%%%%%%%%%%%%%%%%%%%%%%%%%%%%%%%%%%%%%%%%%%%%%%

%\end{document}

%
%%%%%%%%%%%%%%%%%%%%%%%%%%%%%%%%%%%%%%%%%%%%%%%%%%%%%%%%%%%%%%%%%%%%%%%%%%%%%%%
%%%%%%%          P O S T S C R I P T       F I G U R E S 
%%%%%%%   memo:  to include them add epsfig in the \documentstyle
%%%%%%%          and move this part befor \end{document}. 
%%%%%%%          Include the following \newcommand:
%%----------------------------------------------------------------------------
\newcommand{\InsertFigure}[2]{\newpage\begin{center}\mbox{%
\epsfig{bbllx=1.4truecm,bblly=1.3truecm,bburx=19.5truecm,bbury=26.5truecm,%
height=21.4truecm,figure=#1}}\end{center}\vspace*{-1.8truecm}%
\parbox[t]{\hsize}{\small\baselineskip=0.5truecm\hspace*{0.5truecm} #2}}
%----------------------------------------------------------------------------
%%%%%%%%%%%%%%%%%%%%%%%%%%%%%%%%%%%%%%%%%%%%%%%%%%%%%%%%%%%%%%%%%%%%%%%%%%%%%%%
%..............................................................................
\InsertFigure{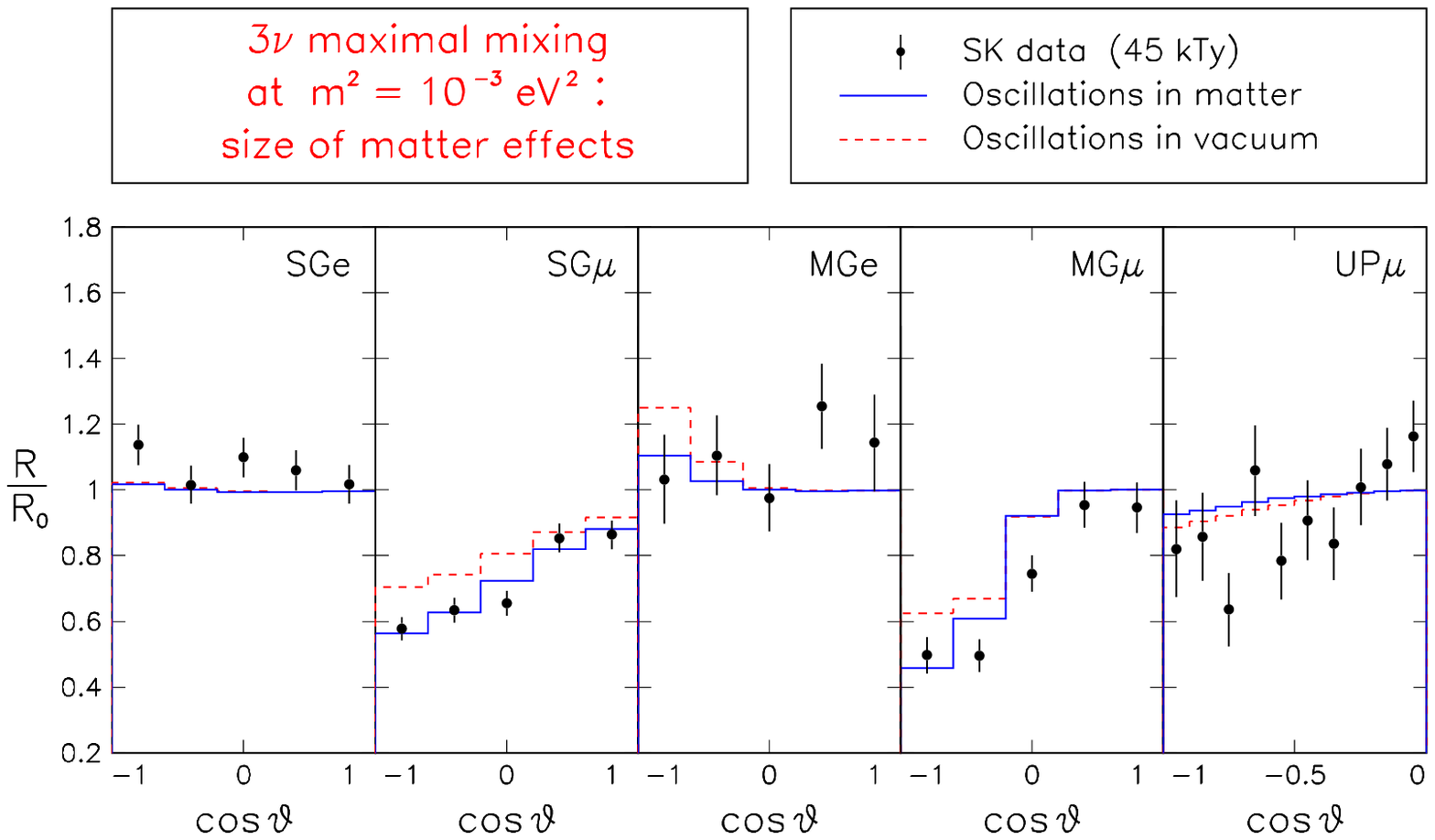}%
{Fig.~1. Threefold maximal mixing scenario at $m^2=10^{-3}$ eV$^2$.
From left to right: zenith distributions of sub-GeV (SG) electrons ($e$)
and muons $(\mu)$, of multi-GeV (MG) $e$ and $\mu$, and of upward going
muons (UP$\mu$), normalized in each bin to the standard expectations
(no oscillation).
Solid line: predictions including matter effects. Dashed lines: predictions
without matter effects. Dots with error bars: Super-Kamiokande data.}
%..............................................................................
\InsertFigure{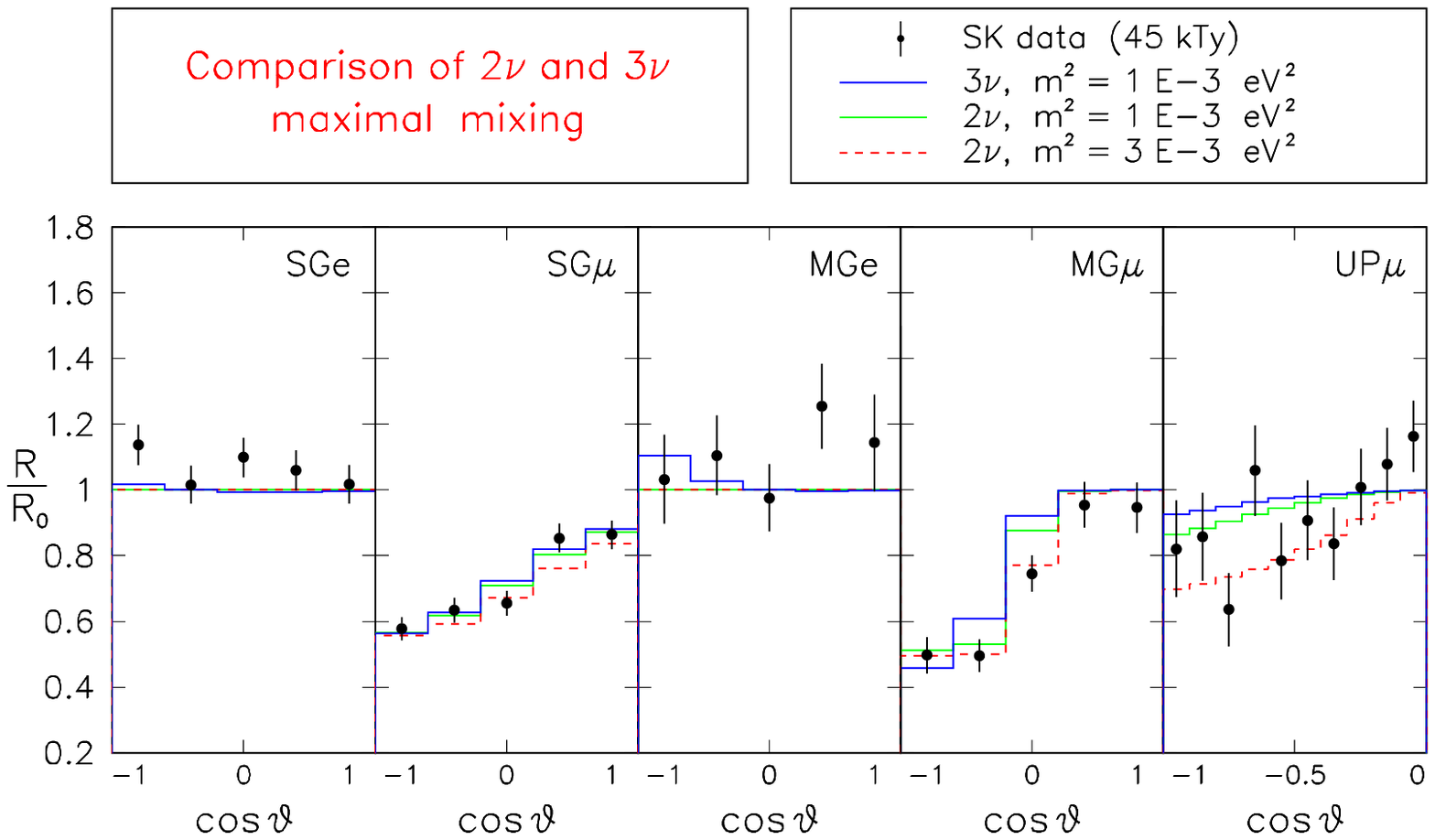}%
{Fig.~2. Comparison of twofold (2$\nu$) and threefold (3$\nu$) maximal
mixing predictions.}
%..............................................................................
\InsertFigure{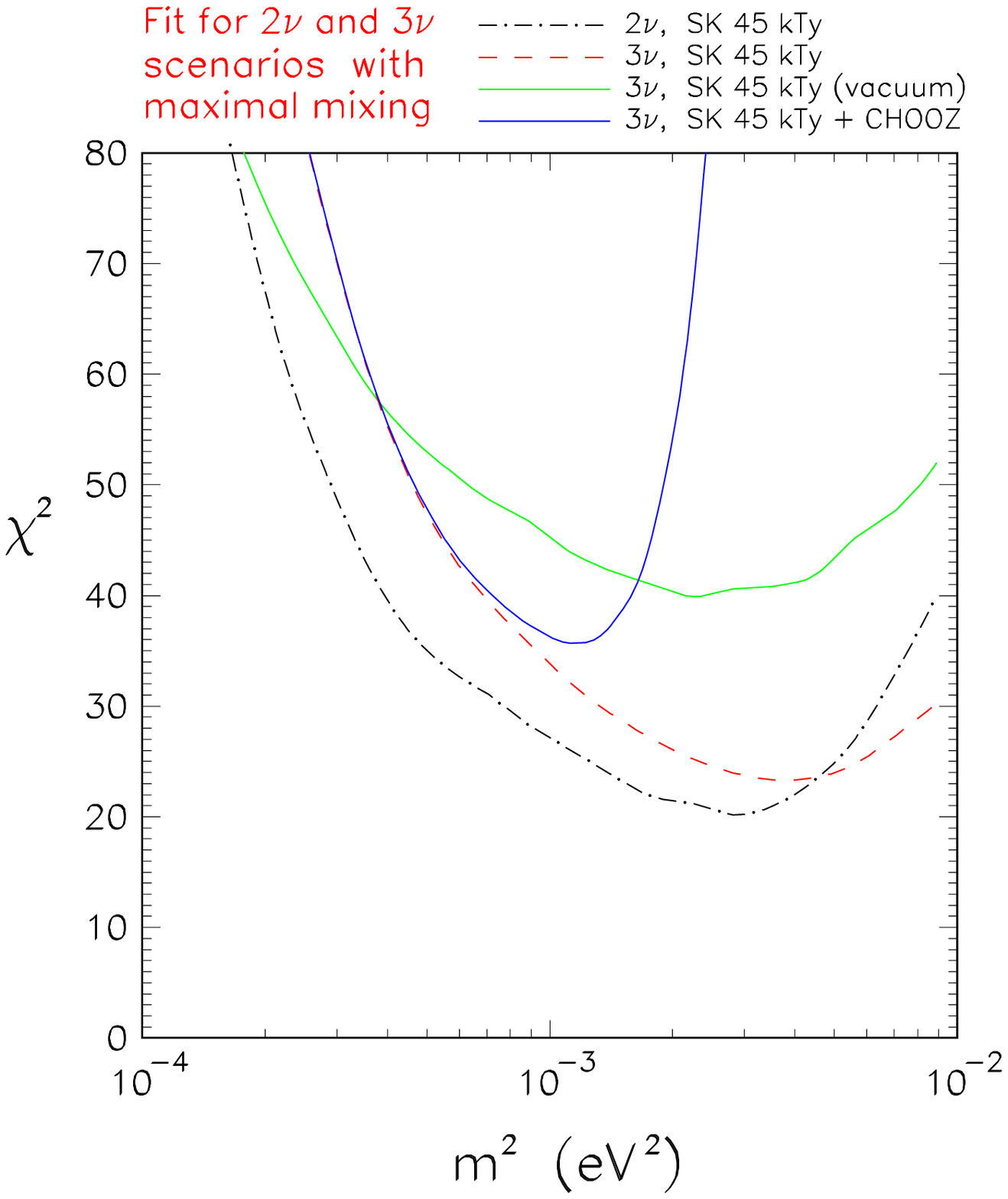}%
{Fig.~3. Fit to twofold and threefold maximal mixing scenarios for variable
$m^2$.}
%..............................................................................
\InsertFigure{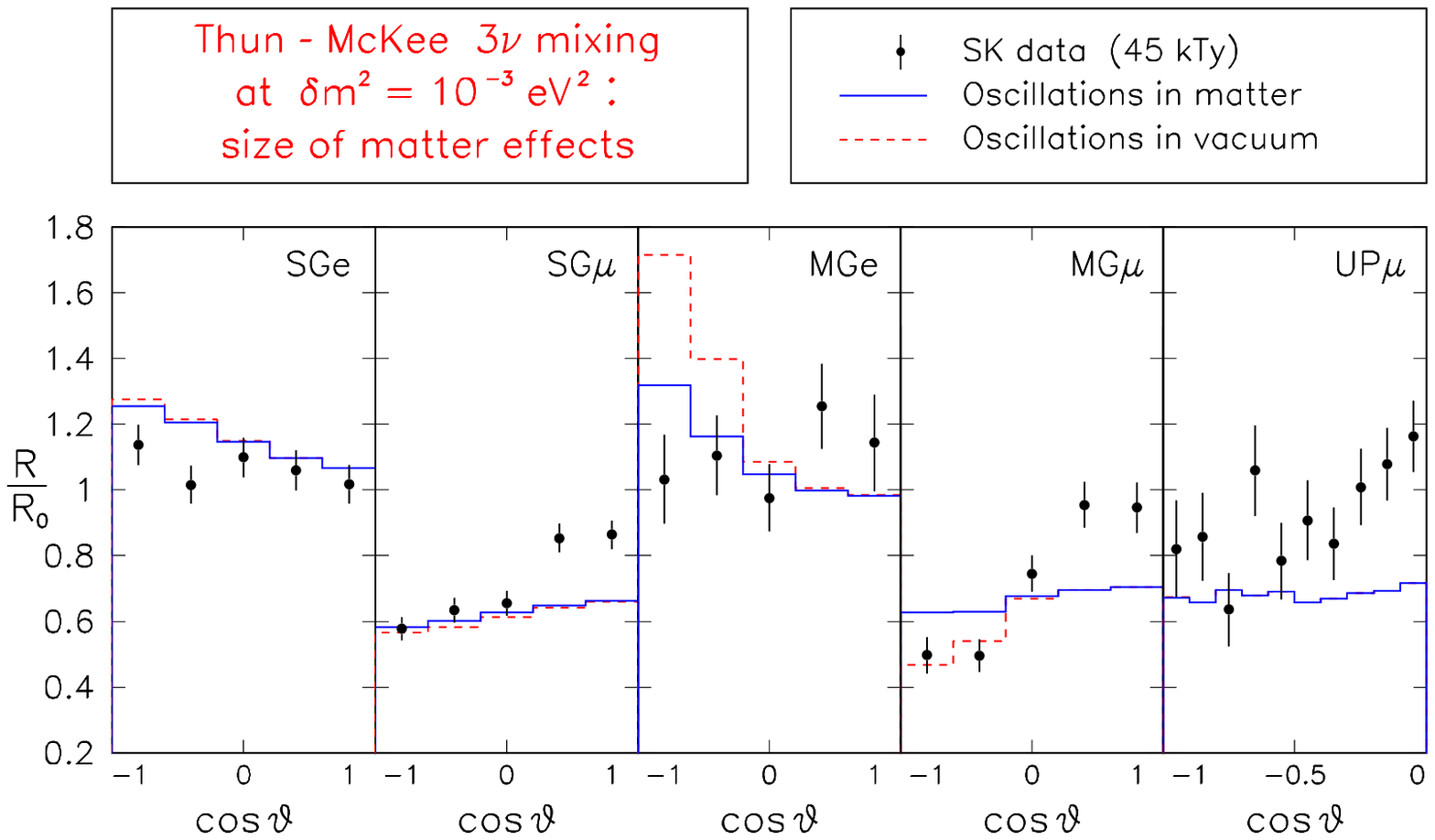}%
{Fig.~4. Thun-McKee $3\nu$ mixing scenario \protect\cite{Th98}
at $\delta m^2=10^{-3}$ eV$^2$, with and without matter effects.}
%..............................................................................
\InsertFigure{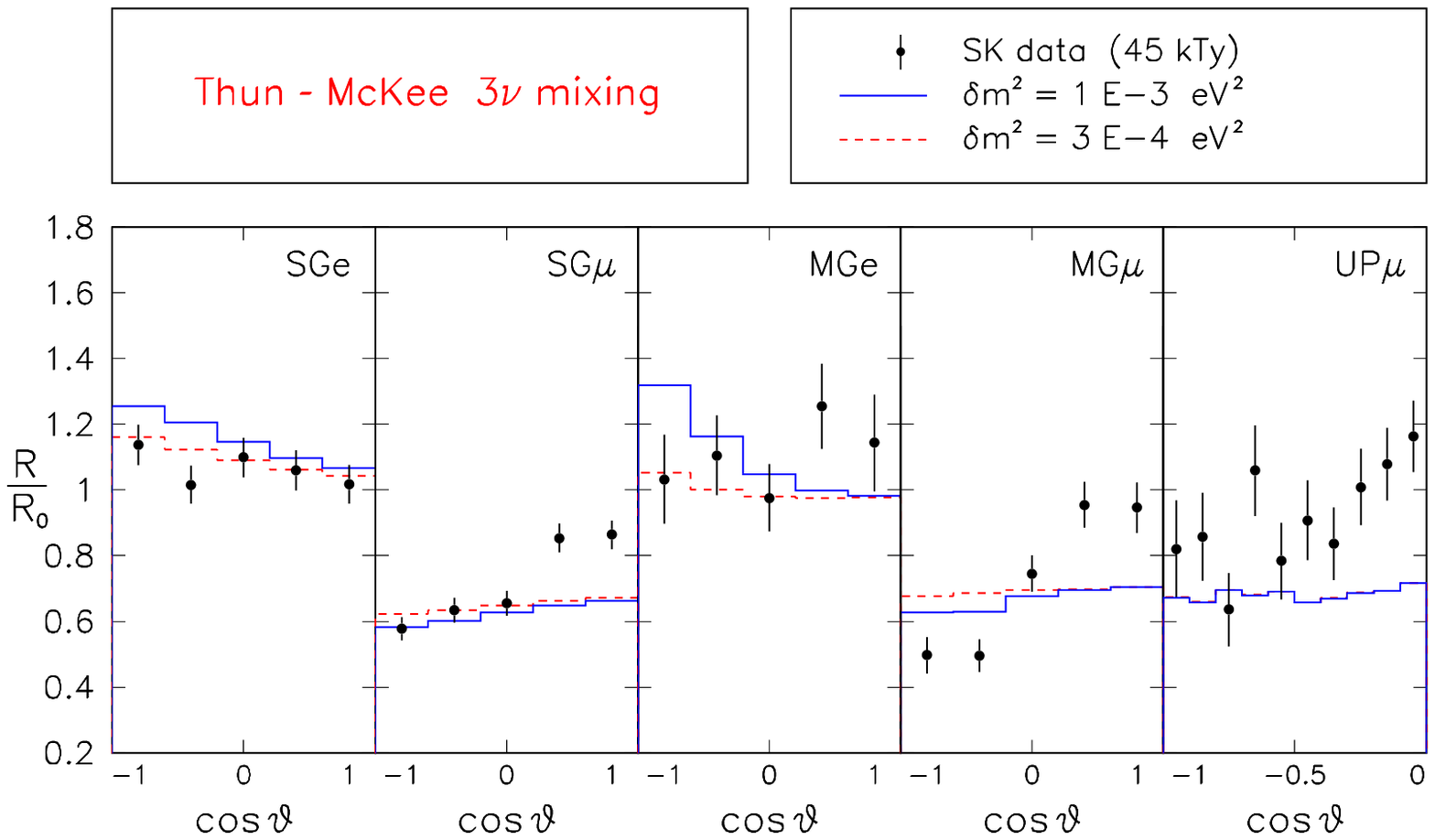}%
{Fig.~5. Thun-McKee $3\nu$ mixing scenarios at $\delta m^2=10^{-3}$
and $3\times 10^{-4}$ eV$^2$ (matter effects included).}

\eject
\end{document}